\documentstyle[12pt,epsf]{article}
\def\href#1#2{{#2}} 
\begin{document}
\title{Loss of quantum coherence through scattering off virtual black holes}
\author{S.W. Hawking$^a$ and Simon F. Ross$^b$ \\ $^a$ Department of
  Applied Mathematics and Theoretical Physics \\ University of
Cambridge, Silver St., Cambridge CB3 9EW \\ {\it
hawking@damtp.cam.ac.uk} \\ $^b$ Department of Physics, University of
California \\ Santa Barbara, CA 93106 \\ {\it
sross@cosmic.physics.ucsb.edu}}
\date{\today \\ DAMTP/R-97/21 \\ UCSB-TH-97-08} 
\maketitle

\begin{abstract}
  In quantum gravity, fields may lose quantum coherence by scattering
  off vacuum fluctuations in which virtual black hole pairs appear and
  disappear. Although it is not possible to properly compute the
  scattering off such fluctuations, we argue that one can get useful
  qualitative results, which provide a guide to the possible effects
  of such scattering, by considering a quantum field on the $C$
  metric, which has the same topology as a virtual black hole pair. We
  study a scalar field on the Lorentzian $C$ metric background, with
  the scalar field in the analytically-continued Euclidean vacuum
  state. We find that there are a finite number of particles at
  infinity in this state, contrary to recent claims made by Yi. Thus,
  this state is not determined by data at infinity, and there is loss
  of quantum coherence in this semi-classical calculation.
\end{abstract}
 
\pagebreak

\section{Introduction}

The possible loss of quantum coherence is one of the most exciting
topics in quantum gravity. Recent work on D-branes has encouraged
those that believe that the evaporation of black holes is a unitary
process without loss of quantum coherence. It has been shown that
collections of strings attached to D-branes with the same mass and
gauge charges as nearly extreme black holes have a number of internal
states that is the same function of the mass and gauge charges as
$e^{A/4G}$, where $A $ is the area of the horizon of the black hole
\cite{strom:D-ent,hor:stringc,mald:D-ent}. They also seem to radiate
various types of scalar particles \cite{das:D-rad1,mald:D-rad} at the
same rate as the corresponding black holes. However, the D-brane
calculations are valid only for weak coupling, at which string loops
can be neglected. But at these weak couplings, the D-branes are
definitely not black holes: there are no horizons, and the topology of
spacetime is that of flat space. One can foliate such a spacetime with
a family of non-intersecting surfaces of constant time. One can then
evolve forward in time with the Hamiltonian and get a unitary
transformation from the initial state to the final state. A unitary
transformation would be a one to one mapping from the initial Hilbert
space to the final Hilbert space.  This would imply that there was no
loss of information or quantum coherence.

To get something that corresponds to a black hole, one has to increase
the string coupling constant until it becomes strong. This means that
string loops can no longer be neglected. However, it is argued that
for gauge charges that correspond to extreme, or near extreme black
holes, the number of internal states will be protected by
non-renormalization theorems, and will remain the same. It is argued
that there's no sign of a discontinuity as one increases the coupling,
and therefore that the evolution should remain unitary. However,
there's a very definite discontinuity when event horizons form: the
Euclidean topology of spacetime will change from that of flat space,
to something non-trivial. The change in topology will mean that any
vector field that agrees with time translations at infinity, will
necessarily have zeroes in the interior of the spacetime. In turn,
this will mean that one cannot foliate spacetime with a family of time
surfaces. If one tries, the surfaces will intersect at the zeroes of
the vector field.  One therefore cannot use the Hamiltonian to get a
unitary evolution from an initial state to a final state. But if the
evolution is not unitary, there will be loss of quantum coherence. An
initial state that is a pure quantum state can evolve to a quantum
state that is mixed.  Another way of saying this is that the
superscattering operator that maps initial density matrices to final
density matrices will not factorise into the product of an $S$ matrix
and its adjoint \cite{hawk:unpred}.  This will happen because the
zeroes of the time translation vector field indicate that there will
be horizons in the Lorentzian section.  Quantum states on such a
background are not completely determined by their asymptotic behavior,
which is the necessary and sufficient condition for the
superscattering operator to factorise.

One cannot just ignore topology and pretend one is in flat space. The
recent progress in duality in gravitational theories is based on
non-trivial topology. One considers small perturbations about
different vacuums of the product form $M^4 \times B$, and shows that
one gets equivalent Kaluza-Klein theories. But if one can have small
perturbations about product metrics, one should also consider larger
fluctuations that change the topology from the product form.  Indeed,
such non-product topologies are necessary to describe pair creation or
annihilation of solitons like black holes or p-branes.

It is often claimed that supergravity is just a low energy
approximation to the fundamental theory, which is string theory.
However, the recent work on duality seems to be telling us that string
theory, p-branes and supergravity are all on a similar footing. None
of them is the whole picture; instead, they are valid in different,
but overlapping, regions. There may be some fundamental theory from
which they can all be derived as different approximations. Or it may
be that theoretical physics is like a manifold that can't be covered
by a single coordinate patch.  Instead, we may have to use a
collection of apparently different theories that are valid in
different regions, but which agree on the overlaps. After all, we know
from Goedel's theorem that even arithmetic can't be reduced to a
single set of axioms. Why should theoretical physics be different?

Even if there is a single formulation of the underlying fundamental
theory, we don't have it yet. What is called string theory has a good
loop expansion, but it is only perturbation theory about some
background, generally flat space, so it will break down when the
fluctuations become large enough to change the topology. Supergravity,
on the other hand, is better at dealing with topological fluctuations,
but it will probably diverge at some high number of loops. Such
divergences don't mean that supergravity predicts infinite answers. It
is just that it cannot predict beyond a certain degree of accuracy.
But in that, it is no different from perturbative string theory. The
string loop perturbation series almost certainly does not converge,
but is only an asymptotic expansion. This means that higher order loop
corrections get smaller at first. But after a certain order, the loop
corrections will get bigger again. Thus at finite coupling, the string
perturbation series will have only limited accuracy.

We shall take the above as justification for discussing loss of
quantum coherence in terms of general relativity or supergravity,
rather than D-branes and strings.  One might expect that loss of
quantum coherence could occur not only in the evaporation of
macroscopic black holes, but on a microscopic level as well, because
of topological fluctuations in the metric that can be interpreted as
closed loops of virtual black holes \cite{hawk:vbh}. Particles could
fall into these virtual black holes, which would then radiate other
particles. The emitted particles would be in a mixed quantum state
because the presence of the black hole horizons will mean that a
quantum state will not be determined completely by its behavior at
infinity. It is with such loss of coherence through scattering off
virtual black holes that this paper is concerned. Our primary
intention is not to provide a rigorous demonstration that quantum
coherence is lost, but rather to explore the effects that will arise,
assuming that the semi-classical calculations are accurate, and it is
lost.

In $d$ dimensions, a single black hole has a Euclidean section with
topology $S^{d-2}\times R^2$. As has been seen in studies of black
hole pair creation, a real or virtual loop of black holes has
Euclidean topology $ S^{d-2}\times S^2 -\{\rm point\}$, where the
point has been sent to infinity by a conformal transformation. For
simplicity, we shall consider $d =4$, but the treatment for higher $d$
would be similar.

On the manifold $ S^2\times S^2 -\{\rm point\}$ one should consider
Euclidean metrics that are asymptotic to flat space at infinity. Such
metrics can be interpreted as closed loops of virtual black holes.
Because they are off shell, they need not satisfy any field equations.
They will contribute to the path integral, just as off shell loops of
particles contribute to the path integral and produce measurable
effects. The effect that we shall be concerned with for virtual black
holes is loss of quantum coherence. This is a distinctive feature of
such topological fluctuations that distinguishes them from ordinary
unitary scattering, which is produced by fluctuations that do not
change the topology.

One can calculate scattering in an asymptotically Euclidean metric on
$ S^2\times S^2 -\{\rm point\}$. One then weights with $\exp (-I)$ and
integrates over all asymptotically Euclidean metrics. This would give
the full scattering with all quantum corrections. However, one can
neither calculate the scattering in a general metric, nor integrate
over all metrics. Instead, what we shall do in the next two sections
is point out some qualitative features of the scattering in general
metrics, that indicate that quantum coherence is lost. We shall then
illustrate the effects of loss of quantum coherence and obtain an
estimate of their magnitude by calculating the scattering in a
specific metric on $ S^2\times S^2 -\{\rm point\}$, the $C$ metric. It
is sufficient to show that quantum coherence is lost in some metrics
in the path integral, because the integral over other metrics cannot
restore the quantum coherence lost in our examples.

\section{Lorentzian section}

We don't have much intuition for the behavior of Euclidean Green
functions or their effect on scattering.  However, if the Euclidean
metric has a hypersurface orthogonal killing vector, it can be
analytically continued to a real Lorentzian metric, in which it is
much easier to see what is happening. We shall therefore consider
scattering in such metrics.

The Lorentzian section of an asymptotically Euclidean metric which has
topology $S^2~\times~S^2 -\{\rm point\}$ will contain a pair of black
holes that accelerate away from each other and go off to infinity. One
might think that this is not very physical, but it is no different
from a closed loop of a particle like an electron. Closed particle
loops are really defined in Euclidean space. If one analytically
continues them to Minkowski space, one gets a particle anti-particle
pair accelerating away from each other. Any topologically non-trivial
asymptotically Euclidean metric will appear to have solitons
accelerating to infinity in the Lorentzian section, but this does not
mean that there are actual black holes at infinity, any more than
there are runaway electrons and positrons with a virtual electron
loop. One can regard the use of the Lorentzian metric, with its black
holes accelerating to infinity, as just a mathematical trick to
evaluate the scattering on the Euclidean solution.

\begin{figure}
\leavevmode
\centering
\epsfysize=15cm 
\epsfbox{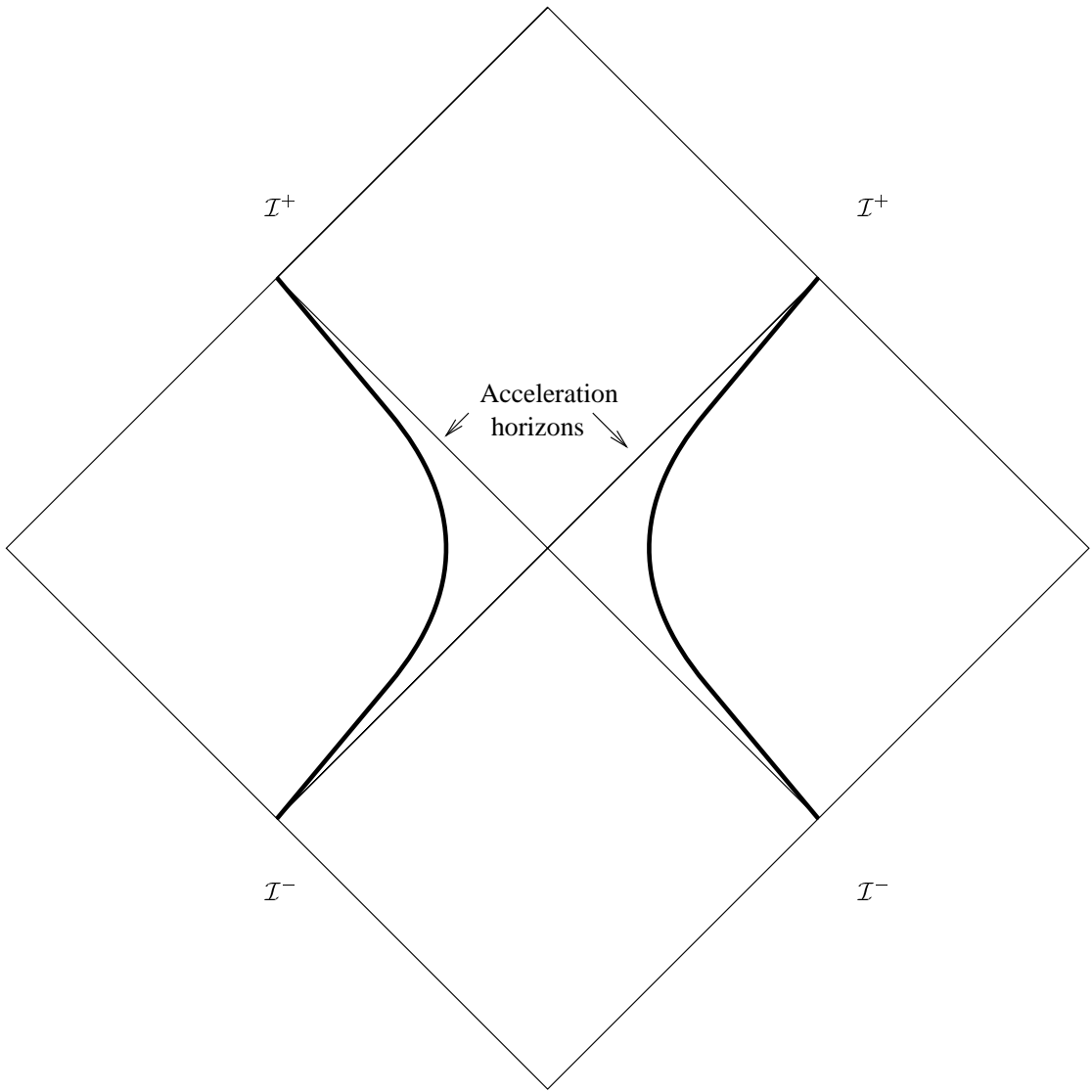}
\caption{ The causal structure of Rindler space, with a pair of
accelerating particles depicted.
\label{rind}}
\end{figure}

To understand the structure of these accelerating black hole metrics,
it is helpful to draw Penrose diagrams.  Start with the Penrose
diagram for Rindler space with the left and right acceleration
horizons, $H_{al}$ and $H_{ar}$, and past and future null infinity,
$\cal I^-$ and $\cal I^+$ (see Figure \ref{rind}).  A uniformly
accelerated particle moves on a world line that goes out to $\cal I^-$
and $\cal I^+$ at the points where they intersect the acceleration
horizons.  One now replaces the accelerating particle and the similar
accelerating particle on the other side with black holes.  Thus, one
replaces the regions of Rindler space to the right and left of the
accelerating world lines with intersecting black hole horizons. It
turns out that the two accelerating black holes are just the two sides
of the same three dimensional wormhole, so one has to identify the two
sides of the Penrose diagram, and the Penrose diagram will look like
the one in Figure \ref{fig1}. At first sight it looks as if one has
lost half of $\cal I^-$ and $\cal I^+$, but that is because this
Penrose diagram applies only on the axis. One can get a better idea of
the causal structure near infinity from Figure \ref{fig2}, in which a
conformal transformation has been used to make $\cal I^+$ into a
cylinder $S^2\times R^1$, with the null generators lying in the $R^1$
direction. The hypersurface orthogonal Killing vector of the Euclidean
metric that allows continuation to a Lorentzian metric will be a boost
Killing vector in the accelerating black hole metric and it will have
two fixed points $q$ and $r$ on $\cal I^+$, lying on generators
$\lambda $ and $\lambda ^{\prime}$ respectively. The past light cones
of $q$ and $r$ minus the generators $\lambda $ and $\lambda ^{\prime}$
form the acceleration horizons.  Thus one can see that nearly every
null geodesic outside the black hole horizons goes out to $\cal I^+$
in the region to the future of both acceleration horizons. The
exceptions are the null geodesics that are exactly in the boost
direction, which intersect the generators $\lambda $ and $\lambda
^{\prime}$. We shall ignore $\lambda $ and $\lambda ^{\prime}$ as a
set of measure zero on $\cal I ^+$, and a number of the statements we
shall make will be valid modulo this set of measure zero.

\section{Quantum state} 

The analytically continued Euclidean Green functions will define a
vacuum state $| 0\rangle_E$ which is the analogue of the so-called
Hartle Hawking state \cite{har:wfn} for a static black hole. The
Euclidean quantum state can be characterized by saying that positive
frequency means positive frequency with respect to the affine
parameters on the horizons. In the accelerating black hole metrics
there are two kinds of horizons, black hole and acceleration. Each
kind of horizon consists of two intersecting null hypersurfaces, which
we shall refer to as left and right, as in Figure \ref{fig1}. In
choosing a Cauchy surface for the spacetime (modulo a set of measure
zero), we break the symmetry between left and right, and choose say
the left acceleration horizon and the right black hole horizon. The
quantum state defined by positive frequency with respect to the affine
parameters on these horizons is the same as the quantum state defined
by the other choice of horizons.

Another Cauchy surface in the future (again modulo a set of measure
zero) is formed by $\cal I ^+$ and the future parts of the black hole
horizons $H_{bl}^+$ and $H_{br}^+$, as in Figure 4.  There is a
natural notion of positive frequency on $\cal I ^+$.  On the black
hole horizons the concept of positive frequency is less well defined.
One could use Rindler time, but in any case, what one observes on
$\cal I ^+$ is independent of the choice of positive frequency on the
black hole horizons.

The quantum state of a field $\phi$ on this background metric will be
determined by data on either of these Cauchy surfaces.  This means
that the Hilbert space $\cal H $ of quantum fields on this background
metric will be isomorphic to the tensor products of the Fock spaces on
their components:
\begin{eqnarray}
\label{hor} {\cal H} &=&{\cal F}_{H_{al}}\otimes {\cal F}_{H_{br}}
\nonumber \\
 &=& {\cal F}_{{\cal I}^+}\otimes {\cal F}_{H_{bl}^+}\otimes
{\cal F}_{H_{br}^+}.
\end{eqnarray} 
The vacuum state defined by the Euclidean Green functions is the
product of the vacuum states of the Fock spaces for the left
acceleration horizon and right black hole horizon;
\begin{equation} \label{state}
| 0\rangle_E= | 0\rangle_{H_{al}} | 0\rangle_{H_{br}}.
\end{equation} 
However, because of frequency mixing, the Euclidean quantum state won't
be the product of the Fock vacuum states on $\cal I ^+$ and the future
black hole horizons. Rather it will be a state containing pairs of
particles.  Both members of the pair may go out to $\cal I ^+$, or both
may fall into the holes, or one go out to $\cal I ^+$ and one fall
in.

Equation (\ref{hor}) shows that quantum field theory on an
accelerating black hole background does not satisfy the asymptotic
completeness condition that the Hilbert space of the quantum fields on
the background is isomorphic to the asymptotic Hilbert space of states
on $\cal I ^+$.  Asymptotic completeness is the necessary and
sufficient condition for scattering of quantum fields on the
background to be unitary \cite{hawk:unpred}. Thus there will be loss of
quantum coherence. What happens is that to calculate the probability
of observing particles at $\cal I ^+$, one has to trace out over all
possibilities on the future black hole horizons. This reduces the
Euclidean quantum state to what appears to be a mixed quantum state
described by a density matrix.

In a recent pair of papers \cite{yi:accbh1,yi:accbh2}, Yi argued that
the Euclidean quantum state in the Ernst metric would contain no
radiation at infinity. The Ernst metric is similar to the metrics we
are considering. However, in the explicit calculation that we carry
out in the $C$ metric, we find that there is indeed radiation at
infinity. What's wrong with Yi's argument? As he was working with the
Ernst metric, which isn't asymptotically flat, he wasn't able to study
the radiation at infinity directly. He therefore assumed that if there
was no radiation on the acceleration horizon, there would be no
radiation at infinity. But if we evolve some state forward from one of
the acceleration horizons to $\cal I^+$, part of the state can fall
into the future black hole horizon. Therefore, there can be a
non-trivial Bogoliubov transformation between the acceleration horizon
and infinity, and Yi's assumption is incorrect.

The Euclidean quantum state $| 0\rangle _E$ will be time symmetric, and
so will contain both incoming and outgoing radiation. Unlike the
Euclidean state for static black holes, there won't be radiation to
infinity at a steady rate for an infinite time. Instead, the radiation
will be peaked around the points $q$ and $r$ where the acceleration
horizons intersect $\cal I ^+$.  The radiation will die off at early
and late times and the total energy radiated will be finite. 

Is this the appropriate quantum state? In the case of a static
black hole, one usually imposes the boundary condition that there is
no incoming radiation on $\cal I ^-$. This means that one has to
subtract the incoming radiation from the Euclidean state to give what
is called the Unruh state. This is singular on the past horizon, but
that doesn't matter, as one normally replaces this region of the
metric with the metric of a collapsing body. The energy for the steady
rate of outgoing radiation comes from a slow decrease of the mass of
the black hole formed by the collapse. However, in the case of a
virtual black hole loop, there is no collapse process to remove the
singularities on the past horizons of the black holes or supply the
energy of the outgoing radiation. Therefore, we should study the
Euclidean vacuum state, in which the energy of the outgoing radiation
is supplied by the incoming radiation on $\cal I ^-$.

Our view therefore is that integrating over gauge equivalent virtual
black hole metrics will cause the amplitude to be zero unless the
energy of the outgoing particle or particles is matched by particles
with the same energy falling in. One might object that one would never
have exactly the combination of incoming particles that corresponded
to the quantum state obtained from the Euclidean green functions.
However, the Euclidean quantum state will appear to be a mixed quantum
state on $\cal I^-$ which contains every possible combination of
incoming particles. One can choose one of these combinations as an
initial pure quantum state that is incident on the virtual black hole
loop. The final quantum state will then be that part of the Euclidean
quantum state on $\cal I ^+$ that has the same energy, momentum and
angular momentum as the incoming state. Because of the trace over the
future black hole horizon states, the final state on $\cal I ^+$ will
be mixed. Such an evolution from pure to mixed states can be described
by a superscattering operator $\$ $ rather than an $S$ matrix
\cite{hawk:unpred}.

The dominant contribution will presumably come from virtual black
hole loops of Planck size. The cross section for a low energy
particle to fall into a Planck size static black hole is very low
unless the particle is spin $0$ or $1/2$ \cite{page:cross}. In the
case of spin $1/2$, the probability of emission will be reduced
because the Fermi-Dirac factor $(\exp (\omega /T)+1)^{-1}$ tends to
$1$ at low $\omega $ while $(\exp (\omega /T)-1)^{-1}$ tends to
$T/\omega $. This suggests the effects of virtual black holes will be
small except for scalar particles. In this paper we shall therefore do
a scattering calculation for scalar particles in the $C$ metric. This
doesn't really qualify as a virtual black hole metric, because it has
conical singularities on the axis, although one can interpret these as
cosmic strings. We study the $C$ metric because it has the same
topological structure as a virtual black hole pair, but it has the
great advantage that one can calculate the scattering, because the
wave equation separates.

\section{$C$ metric}

The charged $C$ metric solution is \cite{kin:Cmetric}
\begin{equation} \label{Cmetric}
  ds^2 = A^{-2} (x-y)^{-2} \left[ G(y) dt^2 - G^{-1}(y) dy^2 +
  G^{-1}(x) dx^2 + G(x) d\varphi^2 \right],
\end{equation}
where
\begin{equation} \label{gxi}
  G(\xi) = (1+r_- A \xi) (1 - \xi^2 - r_+ A \xi^3) = -r_+r_- A^2
  (\xi-\xi_1) (\xi-\xi_2)(\xi-\xi_3)(\xi-\xi_4).
\end{equation}
The gauge potential is
\begin{equation}
  A_\varphi = q (x - \xi_3),
\end{equation}
where $q^2 = r_+ r_-$. We define $m = (r_+ + r_-)/2$.  We
constrain the parameters so that $G(\xi)$ has four roots, which we
label by $\xi_1 \leq \xi_2 < \xi_3 < \xi_4$. To obtain the right
signature, we restrict $x$ to $\xi_3 \leq x \leq \xi_4$, and $y$ to
$-\infty < y \leq x$. The inner black hole horizon lies at $y=\xi_1$,
the outer black hole horizon at $y=\xi_2$, and the acceleration
horizon at $y=\xi_3$. The axis $x=\xi_4$ points towards the other
black hole, and the axis $x = \xi_3$ points towards infinity. Spatial
infinity is at $x=y=\xi_3$, null and timelike infinity at $x=y\neq
\xi_3$. This metric describes a pair of oppositely-charged black holes
accelerating away from each other, although the coordinate system used
in (\ref{Cmetric}) only covers the neighborhood of one of the black holes.

To avoid having a conical singularity
between the two black holes, we choose  
\begin{equation} \label{dphi}
  \Delta \varphi = \frac{4\pi}{|G'(\xi_4)|}.
\end{equation}
This implies that there will be a conical deficit along $x=\xi_3$,
with deficit angle
\begin{equation}
  \delta = 2\pi \left( 1 - \left| \frac{G'(\xi_3)}{G'(\xi_4)} \right|
\right).
\end{equation}
Physically, we imagine that this represents a cosmic string of mass
per unit length $\mu = \delta/8\pi$ along $x=\xi_3$. At large spatial
distances, that is, as $x, y \to \xi_3$, the $C$ metric
(\ref{Cmetric}) reduces to flat space with conical deficit $\delta$ in
accelerated coordinates. The $C$ metric also reduces to flat space if
we set $r_+ =r_- =0$. It reduces to a single static black hole if we
set $A=0$ \cite{dgkt}. The limit $r_+ A \ll 1$ is referred to as the
point-particle limit, as in this limit the black hole is small on the
scale set by the acceleration.  

\begin{figure}
\leavevmode
\centering
\epsfysize=15cm 
\epsfbox{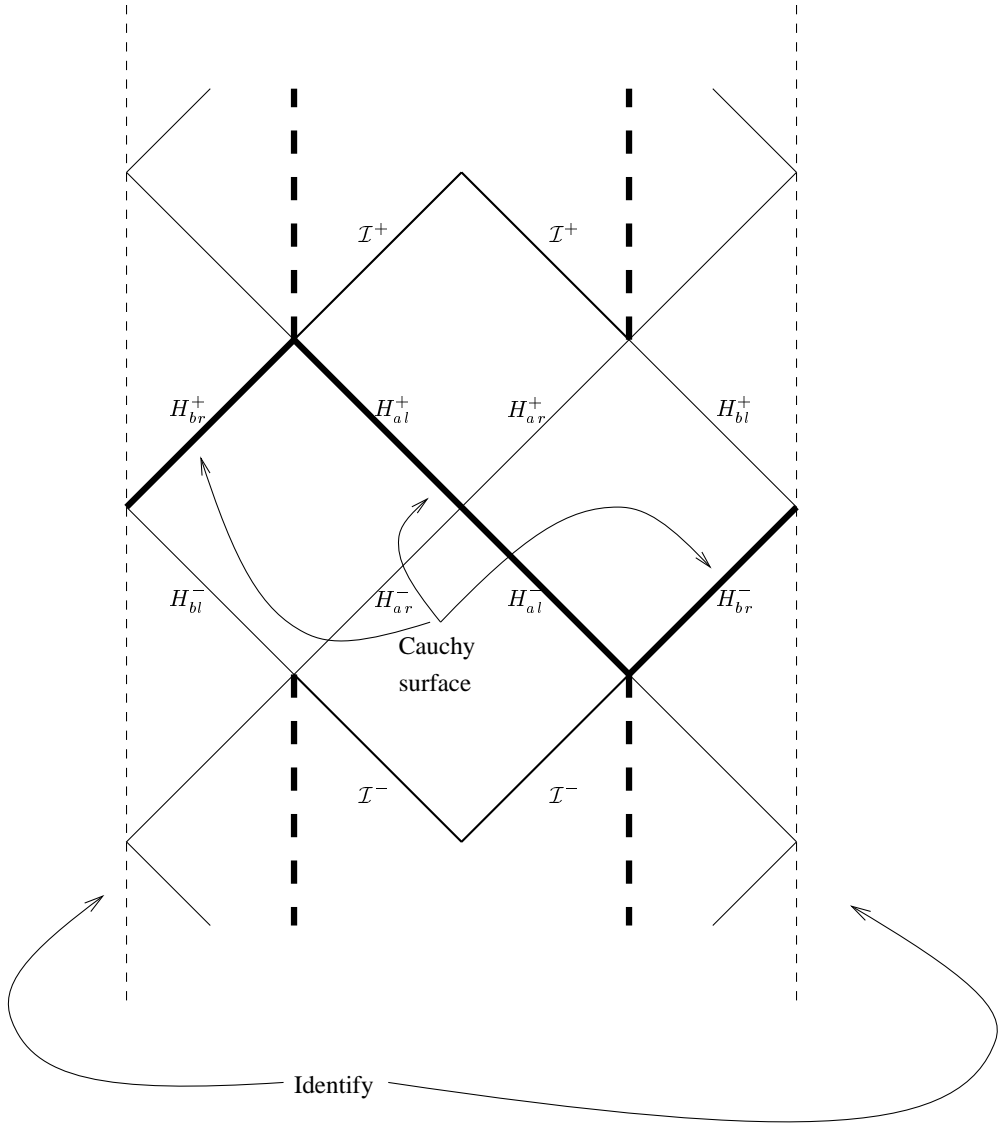}
\caption{ A Penrose diagram for the $C$ metric, neglecting the axis
  $x=\xi_3$. The heavy dashed lines are singularities, and the
  surfaces $\cal I^\pm$ are boundaries of the spacetime. A Cauchy
  surface $\cal C$ for the region outside the inner black hole
  horizons constructed from one black hole horizon and one
  acceleration horizon is shown.
\label{fig1}}
\end{figure}

The $C$ metric was shown to be asymptotically flat in \cite{ash:81}.
This is a considerable advantage over, say, the Ernst metric, as it
means we will have a well-defined notion of $\cal I$, and we can study
the radiation at infinity directly. If we neglect the axis $x =
\xi_3$, all observers will intersect the acceleration horizon before
reaching infinity, and the causal structure of the solution is roughly
speaking given by the Penrose diagram shown in Figure \ref{fig1}.
However, the metric is not spherically symmetric, so this diagram is
not a true picture of the whole spacetime. We will refer to the left
and right acceleration horizons as $H_{al}$ and $H_{ar}$, and to the
left and right outer black hole horizons as $H_{bl}$ and
$H_{br}$. Further, the future and past halves of each horizon will be
denoted by superscripts $\pm$. Hopefully the diagram clarifies the
meaning of this notation.

\begin{figure}
\leavevmode
\centering
\epsfysize=10cm 
\epsfbox{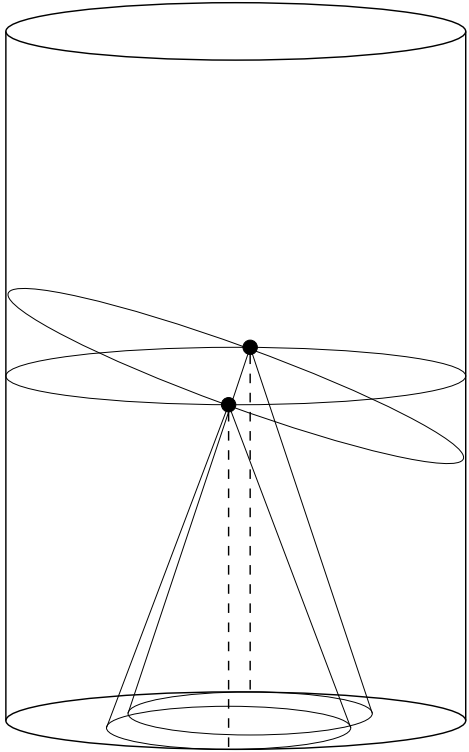}
\caption{The structure of $\cal I^+$ in the conformal
  gauge (\protect\ref{scricm}). The two points are where the black holes
  intersect $\cal I^+$, and their past light cones are the
  acceleration horizons. Two of the $\theta, \eta$ cross-sections are
  pictured. The dashed lines represent the conical deficits in the
  metric (\protect\ref{Cmetric}); they are not part of $\cal I^+$.
\label{fig2}}
\end{figure}

We will only discuss the behavior at future null infinity. As the
metric is time-symmetric, the discussion of past null infinity will be
identical.  We can conformally compactify the $C$ metric by using a
conformal factor $\Omega = A (x-y)$. The conformally rescaled metric
is
\begin{equation}
\tilde{ds}^2 = \Omega^2 ds^2 = G(y) dt^2 - G^{-1}(y) dy^2 +
  G^{-1}(x) dx^2 + G(x) d\varphi^2.
  \label{conC}
\end{equation}
Null infinity is the surface $\Omega=0$, that is, $x=y$ (more
precisely, its maximal extension; the coordinate system of
(\ref{conC}) misses the generator on which the other black hole
intersects $\cal I^+$ \cite{ash:81}). The induced metric on $\cal I^+$
is
\begin{equation} \label{scrim}
 \tilde{ds}^2_{\cal I} = G(y) (dt^2 + d\varphi^2).
\end{equation}
Note that, at null infinity, $t$ is a spatial coordinate. The normal
to $\cal I^+$ is
\begin{equation} \label{norm}
  n^a = \tilde{\nabla}^a \Omega = 2A G(y) \partial_y.
\end{equation}
We see that $t$ and $\varphi$ are constant along the orbits of $n^a$,
which are the generators of $\cal I^+$, so they are good coordinates
on the manifold of orbits of $\cal I^+$. It is convenient to define
new coordinates $\theta, \eta$ where
\begin{equation}
\frac{d\theta}{\sin \theta} = \frac{|G'(\xi_4)|}{2} dt, \qquad \eta =
\frac{|G'(\xi_4)|}{2} \varphi, 
  \label{theta}
\end{equation}
(so $\Delta \eta = 2\pi$). We also make a further conformal rescaling
with a conformal factor $\Omega' = |G'(\xi_4)| \sin \theta/ 2
G^{1/2}(y)$, so that
\begin{equation}
\check{ds}^2_{\cal I} = \Omega'^2 \tilde{ds}^2_{\cal I} = d\theta^2
+ \sin^2 \theta d \eta^2.
  \label{scricm}
\end{equation}
In this conformal gauge, an affine parameter along the generators of
$\cal I^+$ is 
\begin{equation}
\tilde{r} = \frac{|G'(\xi_4)| \sin \theta}{4A} \int \frac{dy}{G(y)^{3/2}}. 
  \label{affine}
\end{equation}
It is also useful to define another coordinate 
\begin{equation} \label{r}
  r = \int \frac{dy}{G(y)^{3/2}},
\end{equation}
which labels the $\theta,\eta$ cross-sections. The structure of $\cal
I^+$ in the conformal gauge (\ref{scricm}) is depicted in Figure
\ref{fig2}. In this conformal gauge, $\cal I^+$ is divergence-free,
and $\theta, \eta$ are coordinates on the manifold of generators of
$\cal I^+$, so we can see that $\cal I^+$ has topology $S^2 \times R$.

We can obtain the Euclidean section of the $C$ metric by setting $t =
i\tau$ in (\ref{Cmetric}). To make the Euclidean metric positive
definite, we need to restrict the range of $y$ to $\xi_2 \leq y \leq
\xi_3$. There are then potentially conical singularities at $y=\xi_2$
and $y=\xi_3$, which have to be eliminated.  We can avoid having a
conical singularity at $y=\xi_3$ by taking $\tau$ to be periodic with
period 
\begin{equation}
  \Delta \tau =\beta = \frac{4\pi}{G'(\xi_3)}.
\end{equation}
In this paper, we assume the black holes are non-extreme, that is,
$\xi_1 < \xi_2$. We can then only avoid having a conical singularity at
$y=\xi_2$ by taking the two horizons to have the same temperature, so
that both conical singularities can be removed by the same choice of
$\Delta \tau$. This implies 
\begin{equation} \label{equaltemp}
  \xi_2 - \xi_1 = \xi_4 - \xi_3.
\end{equation}
The Euclidean section has topology $S^2 \times S^2 - \{pt\}$. This
Euclidean section can be used to study the pair creation of black
holes by breaking cosmic strings
\cite{hawk:strings,greg:strings,eard:cstrings}. However, we want to
use it simply to determine the appropriate vacuum state on the
Lorentzian section. Since the black hole and acceleration horizon have
the same temperature on the Euclidean section, the analytic
continuation will give Green's functions which are thermal with
temperature $1/\beta$ with respect to the time parameter $t$ in the
Lorentzian section.

The region of the spacetime outside the inner horizon of the black
holes is globally hyperbolic. Consider a Cauchy surface for this
region which is made up of one black hole horizon and one acceleration
horizon (say the left acceleration horizon and the right black hole
horizon), as pictured in Figure \ref{fig1}. As explained earlier, the
Hilbert space is isomorphic to the tensor product of the Fock spaces
on the two horizons (\ref{hor}). Positive frequency on the Fock
spaces is defined with respect to the affine parameter along the
horizon. The state we wish to study is the analytically-continued
Euclidean vacuum state $|0\rangle_E$ given in (\ref{state}).

In the next section, we will describe the solution of the scalar wave
equation on the $C$ metric background. We then use this to calculate
the Bogoliubov coefficients in the subsequent section. 

\section{Scalar Wave Equation}

We consider a minimally-coupled massless neutral scalar field, so the
wave equation is just $\Box \phi =0$. One of the great advantages of
considering the $C$ metric is that this equation separates. It is easy
to see this if we observe that the $C$ metric is a solution of the
vacuum Einstein-Maxwell equations, and hence $R=0$. The minimally
coupled equation above is therefore equivalent to the
conformally-invariant equation $\Box \phi - \frac{1}{6}R \phi =0$. But
in solving this latter equation, we are free to make conformal
transformations. In particular, we can transform to the conformal
gauge (\ref{conC}), in which this equation takes the form
\begin{equation}
\frac{1}{G(y)} \partial_t \partial_t \tilde{\phi}  - \partial_y [G(y)
\partial_y \tilde{\phi}] + \partial_x [G(x) \partial_x \tilde{\phi}] +
\frac{1}{G(x)}\partial_\varphi \partial_\varphi \tilde{\phi} +
\frac{1}{6}[\partial_x^2G(x) - \partial_y^2G(y)] \tilde{\phi} = 0,
  \label{cweq}
\end{equation}
where because of the conformal rescaling, $\tilde{\phi} =
\phi/A(x-y)$. Thus we see that if we use the ansatz
\begin{equation}
\phi = A(x-y) e^{i\omega t} e^{i m\varphi} \nu(x) \gamma(y), 
  \label{ans}
\end{equation}
then we get two second-order ODEs for $\nu(x)$ and $\gamma(y)$,
\begin{equation}
\partial_x [G(x) \partial_x \nu(x)] - \frac{1}{G(x)}m^2 \nu(x)+
[\frac{1}{6} \partial_x^2G(x) + D] \nu(x) = 0 
  \label{xeq}
\end{equation}
and
\begin{equation}
\partial_y [G(y) \partial_y \gamma(y)] + \frac{1}{G(y)}\omega^2
\gamma(y)+  [\frac{1}{6} \partial_y^2G(y) + D] \gamma(y) = 0,
  \label{yeq}
\end{equation}
where $D$ is a separation constant, and $G(\xi)$ is given in
(\ref{gxi}). Note that $\varphi$ is a periodic coordinate
with period $4\pi/|G'(\xi_4)|$. Thus $m = m_0 |G'(\xi_4)|/2$, where
$m_0$ is an integer. We assume, without loss of generality,
that it is positive. 

One way to rewrite these equations that offers some further insight is
to define new coordinates
\begin{equation} \label{zchi}
z = \int \frac{dy}{G(y)}, \qquad \chi = \int \frac{dx}{G(x)},
\end{equation}
which have the advantage that $\partial_z = G(y) \partial_y,
\partial_\chi = G(x) \partial_x$. Note that the integral for $z$ in
(\ref{zchi}) diverges as we approach a horizon, as $G(y) \to 0$ at the
horizons. Thus, $-\infty < z <\infty$ only covers the region between
two of the horizons; similarly, $\xi_3 < x < \xi_4$ is mapped to
$-\infty < \chi < \infty$. We can write (\ref{xeq},\ref{yeq}) as
\begin{equation}
\partial_\chi^2 \nu(x(\chi)) - m^2 \nu(x(\chi)) + V_{eff}(\chi)
\nu(x(\chi)) = 0, 
  \label{chieq}
\end{equation}
\begin{equation}
\partial_z^2 \gamma(y(z)) + \omega^2 \gamma(y(z)) + V_{eff}(z)
\gamma(y(z)) = 0.
  \label{zeq}
\end{equation}
That is, (\ref{yeq}) reduces to the one-dimensional wave equation with
effective potential $V_{eff}(z)$, which is given by
\begin{equation}
  V_{eff}(z) = G(y(z))[\frac{1}{6} \partial_y^2G(y(z)) + D].
\end{equation}
There is a similar expression for $V_{eff}(\chi)$. It is not possible
to invert (\ref{zchi}) to obtain $y(z)$ explicitly, but we can make
some observations. Near the horizons, $G(y) \to 0$, and thus the
effective potential becomes unimportant, so $\gamma(y) \sim e^{\pm i
\omega z}$.  Similarly, near $x = \xi_3, \xi_4$, $\nu(x) \sim e^{\pm m
\chi}$.  Obviously, for physically-interesting solutions, we must have
$\nu(x) \sim e^{-m|\chi|}$ as $\chi \to \pm \infty$.

\begin{figure}
\leavevmode
\centering
\epsfysize=15cm 
\epsfbox{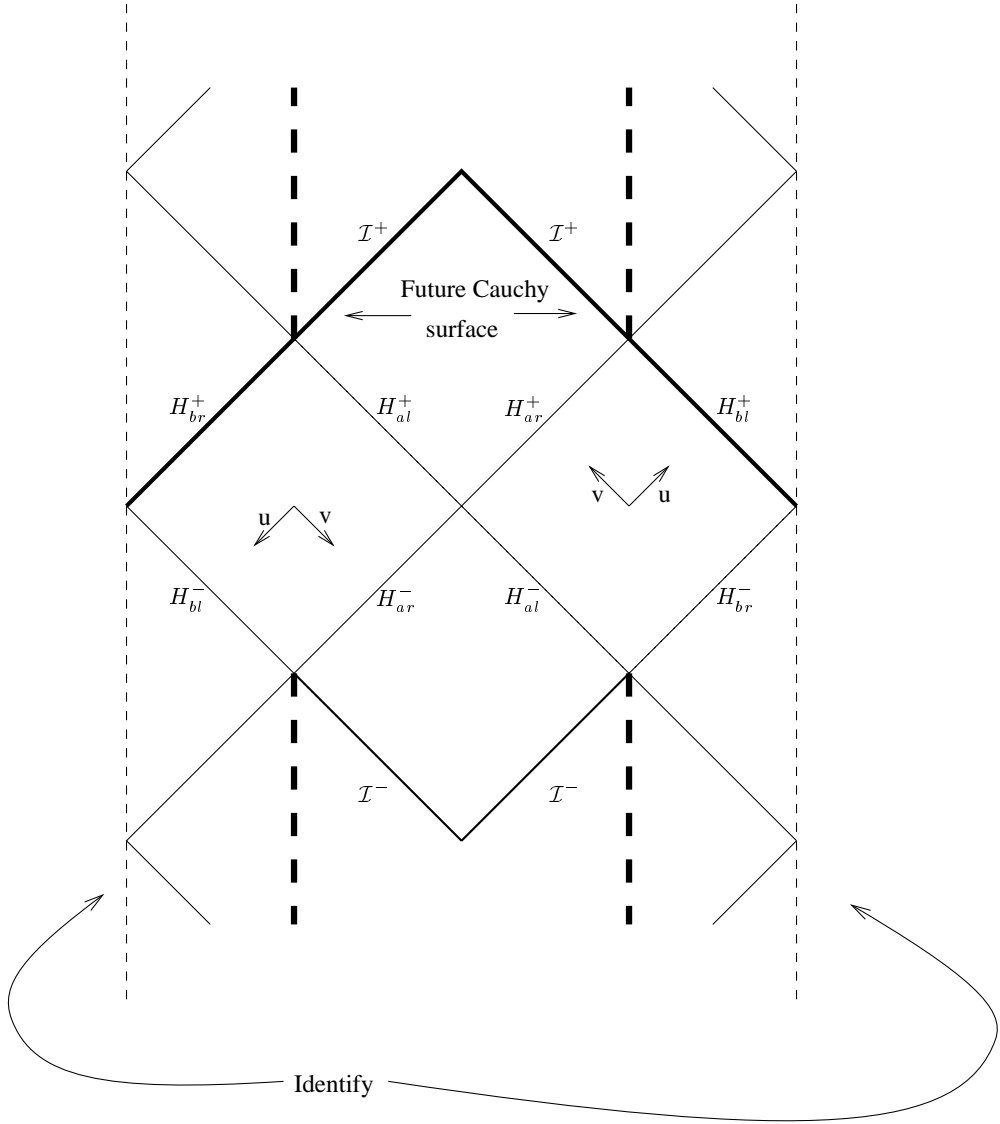}
\caption{A Cauchy surface $\tilde{\cal C}$ for the region outside the
  inner black hole horizons constructed from $\cal I^+$ and the future
  halves of the black hole horizons. The Rindlerian coordinates $u,v$
  between the acceleration and outer black hole horizons are also
  shown.
\label{fig3}}
\end{figure}

We can rewrite the metric (\ref{Cmetric}) in terms of these
coordinates:
\begin{equation}
  ds^2 = A^{-2}(x-y)^{-2}[ G(y)(dt^2 - dz^2) + G(x)(d\chi^2 +
  d\varphi^2)],
\end{equation}
where by $x,y$ we mean $x(\chi)$, $y(z)$. This coordinate system
evidently only covers the region between two of the horizons (or
between the acceleration horizon and infinity). That is, there is a
coordinate system like this for each of the diamond-shaped regions in
the Penrose diagram in Figure \ref{fig1}. We will therefore refer to
these as the Rindlerian coordinates. We can now define null
coordinates $u,v = t\pm z$. Since $z$ increases as we go from the
acceleration horizon towards the black hole horizon, the $u$ and $v$
coordinates run as shown in Figure \ref{fig3}. Thus, $u$ is a
(non-affine) parameter along $H_{ar}^{\pm}$ and $H_{br}^{\pm}$, while
$v$ is a (non-affine) parameter along $H_{al}^{\pm}$ and
$H_{bl}^{\pm}$. As is usual for bifurcate Killing horizons, these
parameters are related to the affine parameters $U, V$ on the
acceleration horizon by $u = \frac{1}{\kappa} \ln |U|$, $v =
-\frac{1}{\kappa} \ln |V|$, where $\kappa = G'(\xi_3)/2$ is the common
surface gravity of the two horizons.

These coordinates are useful for specifying boundary conditions near
the black hole and the acceleration horizons, and we will see later
that we can easily write down explicit forms for the
positive-frequency wavefunctions on the horizons in terms of them.
However, as we can't write $V_{eff}$ explicitly as a function of $z$,
we can't solve the differential equations in this form.

If we return to the initial forms (\ref{xeq},\ref{yeq}) for the ODEs,
we find that they can be considerably simplified. In the
simplification, we will exploit the equal-temperature condition
(\ref{equaltemp}), which imposes an additional symmetry on the form of
$G(\xi)$. If we make a coordinate transformation
\begin{equation} \label{hat}
\hat{\xi} = \frac{2}{(\xi_3 - \xi_2)}[\xi - \frac{1}{2} (\xi_3 + \xi_2)],
\end{equation}
then
\begin{equation}
  G(\xi) = -\frac{\psi}{\zeta} (\hat{\xi}^2 - \alpha^2)(\hat{\xi}^2 - 1),
\end{equation} 
where 
\begin{equation}
  \zeta = \frac{8}{r_+ r_- A^2 (\xi_3-\xi_2)^3}, \qquad \alpha =
  \frac{(\xi_4 - \xi_1)}{(\xi_3 - \xi_2)}, \qquad \psi = \frac{1}{2}
  (\xi_3 - \xi_2).
\end{equation}
Note that $\alpha >1$, $\zeta, \psi > 0$, and that
$\partial_{\hat{\xi}} = \psi \partial_\xi$. If $\hat{y}$ and
$\hat{x}$ are defined in terms of $y$ and $x$ following (\ref{hat}),
then the inner black hole horizon is at $\hat{y} = -\alpha$, the
outer black hole horizon is at $\hat{y} = -1$, and the acceleration
horizon is at $\hat{y} = 1$, while the range of $\hat{x}$ is $1
\leq \hat{x} \leq \alpha$.  In terms of these coordinates,
\begin{equation}
  \partial_\xi^2 G(\xi) = -\frac{1}{\zeta \psi} [12 \hat{\xi}^2 - 2(1
  +\alpha^2)],  
\end{equation}
so it is convenient to define
\begin{equation}
  \beta_D^2 = \frac{1}{6}
(1 + \alpha^2) + \frac{D \psi \zeta}{2},
\end{equation}
so that 
\begin{equation}
  \frac{1}{6} \partial_\xi^2 G(\xi) + D = -\frac{2}{\zeta \psi}
  (\hat{\xi}^2 - \beta_D^2). 
\end{equation}
We can now write $z$ explicitly; 
\begin{equation}
  z = \int \frac{dy}{G(y)} = \frac{\zeta}{2(\alpha^2-1)} \left[
  \frac{1}{\alpha} \ln \left| \frac{\alpha + \hat{y}}{\alpha -
    \hat{y}} \right| + \ln \left| \frac{1-\hat{y}}{1+\hat{y}}
\right|\right].  
\end{equation}
We can now see clearly that $z$ diverges at the event horizons
$\hat{y}= -\alpha, \pm1$. We can further see that $z \to -\infty$ as
we approach $\hat{y} = -\alpha, 1$, the inner black hole and
acceleration horizons, and $z \to \infty$ as we approach $\hat{y}=-1$,
the outer black hole horizon. There is a similar explicit expression
for $\chi$, and $\chi \to -\infty$ as we approach $\hat{x} =1$ and
$\chi \to \infty$ as we approach $\hat{x} = \alpha$. The consideration
of the form (\ref{chieq},\ref{zeq}) suggests a further simplifying
transformation. If we set
\begin{equation}
  \hat{\nu}(\hat{x}) = e^{m \chi} \hat{n}(\hat{x})  = \left(
  \frac{\alpha+\hat{x}}{\alpha-\hat{x}} \right)^{\frac{\zeta m}{2
      \alpha (\alpha^2-1)}} \left(
  \frac{\hat{x}-1}{1+\hat{x}} \right)^{\frac{\zeta m}{2
      (\alpha^2-1)}} \hat{n}(\hat{x}) 
\end{equation}
and
\begin{equation}
  \hat{\gamma}(\hat{y}) = e^{-i \omega z} \hat{f}(\hat{y}) = \left(
  \frac{\alpha+\hat{y}}{\alpha-\hat{y}} \right)^{\frac{-i \zeta
  \omega}{2 \alpha (\alpha^2-1)}} \left( \frac{1-\hat{y}}{1+\hat{y}}
  \right)^{\frac{-i \zeta \omega}{2 (\alpha^2-1)}} \hat{f}(\hat{y}),
\end{equation}
then we can finally rewrite (\ref{xeq},\ref{yeq}) as 
\begin{equation}
\partial_{\hat{x}}[(\hat{x}^2 -1)(\hat{x}^2-\alpha^2)
\partial_{\hat{x}} \hat{n}(\hat{x})] -2m\zeta \partial_{\hat{x}}
\hat{n}(\hat{x}) + 
2(\hat{x}^2 - \beta_D^2) \hat{n}(\hat{x}) = 0,
  \label{nxeq}
\end{equation}
\begin{equation}
\partial_{\hat{y}}[(\hat{y}^2 -1)(\hat{y}^2-\alpha^2)
\partial_{\hat{y}} \hat{f}(\hat{y})] +2i\omega \zeta
\partial_{\hat{y}} \hat{f}(\hat{y}) + 
2(\hat{y}^2 - \beta_D^2) \hat{f}(\hat{y}) = 0.
  \label{fyeq}
\end{equation}
This is the simplest form in which we can write these equations. 

We have been able to simplify the form of the wave equation
considerably. However, (\ref{nxeq},\ref{fyeq}) still have five regular
singular points, at $\hat{\xi} = \pm 1, \pm \alpha, \infty$, so they
can't be solved exactly. We will therefore need to use some further
simplifying assumption in solving the wave equation. There is only one
dimensionless parameter in the metric, $r_+A$, as the
equal-temperature condition fixes $r_-A$ as a function of
$r_+A$. Therefore we are driven to consider the point-particle limit
$r_+ A \ll 1$.  In this limit, $\alpha \approx 1 + 4r_+ A$, and $\zeta
\approx 8 r_+ A \approx 2 (\alpha-1)$.  For reasons of convenience, we
will use $(\alpha-1)$ as the small parameter.

\section{Bogoliubov Transformations}
\label{bog}

Having laid the groundwork, we can now define and evaluate the
Bogoliubov coefficients. We can write the field operator $\phi$ in
terms of annihilation and creation operators on the Hilbert spaces
associated with the black hole and acceleration horizons:
\begin{equation} \label{ops1}
\phi = A(x-y) \Sigma_{lm}
\int d\omega ( f^{b}_{\omega lm} b^{b}_{\omega lm} +
\bar{f}^{b}_{\omega lm} b^{b \dagger}_{\omega lm} +
f^{a}_{\omega lm} b^{a}_{\omega lm} + 
\bar{f}^{a}_{\omega lm} b^{a \dagger}_{\omega lm}),
\end{equation}
where $f^{b}_{\omega lm}, f^{a}_{\omega lm}$ are sets of positive
frequency modes which have non-zero support on the black hole and
acceleration horizons respectively, $b^{b}_{\omega lm}, b^{a}_{\omega
lm}$ are the particle annihilation operators, and $b^{b
\dagger}_{\omega lm}, b^{a \dagger}_{\omega lm}$ are the particle
creation operators. Here, positive frequency means with respect to the
affine parameters $U,V$ on the horizons.

Following \cite{wald:qft}, we see that a suitable set of positive
frequency states on the black hole horizon is
\begin{equation} \label{bhp1}
 f^{b}_{\omega lm} = \frac{ N}{|1-e^{-2\pi 
\omega/\kappa}|^{1/2}} e^{im\varphi} \nu_{lm}(x)[ g_\omega^- +
e^{-\pi \omega/\kappa} g_\omega^+],
\end{equation}
where $\nu_{lm}$ is a solution of (\ref{xeq}) with $D$ given by
$\beta_D = 1 + 2l(l+1)$, and $g_\omega^\pm$ are functions which are
non-zero on the future and past parts of the black hole horizon
respectively, and which are positive frequency with respect to the
Rindler parameter, that is, $g_\omega^\pm = e^{-i\omega u}$. We know
already that only a discrete set of values for $m$ are allowed, and we
will see below that the same is true for $l$. We wish to normalize the
modes so that $(f^{b}_{\omega lm}, f^{b}_{\omega' l' m'}) =
\delta_{mm'} \delta_{ll'} \delta(\omega-\omega')$, which implies
$|N|^2 = 1/(4\pi|\omega| \Delta \varphi)$. Note that although the
positive-frequency solutions are labeled by a frequency $\omega$,
they do not have a single frequency with respect to $U$, and the
solutions are still wholly positive frequency with respect to $U$ when
$\omega$ is negative. For this to be a complete set of
positive-frequency solutions, we must allow $\omega$ to run over
$-\infty < \omega < \infty$. One can write down a similar set of
positive frequency solutions on the acceleration horizon.

In appendix \ref{ang}, we consider (\ref{nxeq}) with $(\alpha-1) \ll
1$, and we learn that, as we might have expected, there is a
restriction on the form of the data on the black hole horizon.  If we
write $l = l_0 + O(\alpha-1)$, then the solutions $\nu_{lm}(x)$ will
only be regular at both of the axes $x=\xi_3, x=\xi_4$ if $l_0$ is an
integer and $l_0 \geq m_0$, where $m_0$ is the integer appearing in
$m$. In the point-particle limit, the $x,\varphi$ section approaches
spherical symmetry, so $l_0$ is the usual total angular momentum
quantum number, while $m_0$ is the angular momentum with respect to
the axis along which the black holes are accelerating.

We can also write the field operator in terms of modes which are
positive frequency at infinity:
\begin{equation} \label{ops2}
\phi = A(x-y) \int d\omega\, d\theta_0\, d\eta_0\, ( p_\omega a_\omega
+ \bar{p}_\omega a^\dagger_\omega  
+ q_\omega c_\omega + \bar{q}_\omega c^\dagger_\omega),
\end{equation}
where $p_\omega$ are a set of modes with non-zero support on ${\cal
I}^+$ which are positive frequency with respect to $\tilde{r}$, and
$a_\omega, a^\dagger_\omega$ are the corresponding annihilation and
creation operators. The modes $q_\omega$ have non-zero support on the
future black hole horizon, and $c_\omega, c^\dagger_\omega$ are the
corresponding annihilation and creation operators. We won't bother to
define these latter modes, as their form is irrelevant to the
calculation of particle production on $\cal I^+$.

Following \cite{hawk:bub2}, we choose the positive frequency modes
$p_\omega$ to have the form
\begin{equation}
p_\omega = \frac{e^{-i \tilde{\omega} \tilde{r}}}{\sqrt{2\pi}
\tilde{\omega} \sin 
\theta_0} \delta(\theta - \theta_0) \delta(\eta - \eta_0) =
\frac{e^{-i \omega r}|G'(\xi_4)|}{\sqrt{2\pi} 4A \omega}
\delta(\theta - \theta_0) \delta(\eta - \eta_0)
\end{equation}
on $\cal I^+$, in the conformal gauge where the metric on $\cal I^+$
has the form (\ref{scricm}). We define $\omega = \tilde{\omega}
|G'(\xi_4)| \sin \theta_0/ 4A $. Each mode is thus non-zero on one
generator of $\cal I^+$, labeled by $\theta_0,\eta_0$, and has
frequency $\tilde{\omega}$ with respect to the affine parameter along
that generator. The complete set of positive frequency modes is given
by $0 \leq \omega < \infty$. They are normalized so that
$(p_\omega,p_\omega') = 2 \tilde{\omega} \delta^3(\vec{k} -
\vec{k}')$, where $\vec{k}$ is the three-momentum, and points in the
direction $(\theta_0,\eta_0)$.

Since both sets of modes are complete bases for the space of solutions
of the wave equation, we can write one in terms of the other. That is,
\begin{equation} \label{bog1}
f^{b}_{\omega' lm}=  \int d\tilde{\omega}\, d\theta_0\, d\eta_0\,
(\alpha^{b}_{\omega \omega' lm} 
p_\omega + \beta^{b}_{\omega \omega' lm}
\bar{p}_\omega + \mbox{terms involving }q_\omega),
\end{equation}
and similarly for $f^{a}_{\omega' lm}$. If we substitute these
expansion into (\ref{ops1}), and require consistency with
(\ref{ops2}), then we find that
\begin{equation} \label{bog2}
a_\omega = \Sigma_{lm} \int d\omega' (\alpha^{b}_{\omega \omega' lm}
b^{b}_{\omega' lm}  +\bar{\beta}^{b}_{\omega \omega' lm}
b^{b\dagger}_{\omega' lm} + \alpha^{a}_{\omega \omega' lm}
b^{a}_{\omega' lm} + \bar{\beta}^{a}_{\omega \omega' lm}
b^{a\dagger}_{\omega' lm}).
\end{equation}
The quantities $\alpha^{b}_{\omega \omega' lm}, \beta^{b}_{\omega
\omega' lm}, \alpha^{a}_{\omega \omega' lm}, \beta^{a}_{\omega \omega'
lm}$ are called the Bogoliubov coefficients. Since we know how the
annihilation and creation operators which were defined on the horizons
act on $|0\rangle_E$, to determine how the annihilation and creation
operators defined at infinity act on $|0\rangle_E$, we just need to
compute these coefficients.

The operator we are most interested in is the number operator
$N_\omega = a^\dagger_\omega a_\omega$, which gives the number of
particles in the mode $p_\omega$. In the state $|0\rangle_E$,
\begin{eqnarray} \label{numop}
\langle 0|N_\omega | 0 \rangle_E &=& \Sigma_{lml'm'} \int d \omega' d
\omega'' ( 
\beta^{b}_{\omega \omega' lm} \bar{\beta}^{b}_{\omega \omega'' l'm'}
\langle 0| b^{b}_{\omega' lm} b^{b\dagger}_{\omega'' l'm'}  | 0
\rangle_E + \ldots ) \\ \nonumber 
&=& \Sigma_{lm} \int_{-\infty}^{\infty} d\omega'
(|\beta^{b}_{\omega \omega' lm}|^2 +|\beta^{a}_{\omega
\omega' lm}|^2), 
\end{eqnarray}
where we have expanded $a_\omega$ by (\ref{bog2}), and in the second
line we have used the canonical commutation relations and the fact
that $b^{b}_{\omega' lm} |0\rangle_E =0, b^{a}_{\omega' lm}
|0\rangle_E =0$.

We should now calculate the Bogoliubov coefficients $\beta^{b}_{\omega
  \omega' lm}$ and $\beta^{a}_{\omega \omega' lm}$. However, it turns
out to be quite difficult to calculate the latter coefficient.
Therefore, we wish to argue that it is sufficient to calculate the
contribution from the Bogoliubov coefficient associated with the black
hole horizon $\beta^{b}_{\omega \omega' lm}$; the other contribution
should be similar. 

We broke the symmetry between the left and right horizons when we
defined the Euclidean vacuum state, by defining it to be the product
of the vacua of the Fock spaces for the left acceleration horizon and
right black hole horizon. However, the vacuum state is in fact
symmetric under left-right interchange. That is, it is also equal to
the product of the vacua of the Fock spaces for the right acceleration
horizon and left black hole horizon. Take the vacuum state on $\cal C$
and evolve it forward through the right Rindler diamond, from
$H_{al}^{-}$ and $H_{br}^{-}$ to $H_{ar}^{+}$ and $H_{bl}^{+}$. There
will then be correlations between $H_{br}^{+}$ and $H_{ar}^{+}$, due
to the correlations between the two halves of the black hole horizon
in the Cauchy surface $\cal C$. Further, there are no correlations
between $H_{br}^{+}$ and $H_{al}^+$, because on $\cal C$, the state
has no correlations between the black hole and acceleration horizons.
Since the state is left-right symmetric, the correlations between the
two halves of the acceleration horizon in the Cauchy surface $\cal C$
can therefore only give rise to correlations between $H_{bl}^{+}$ and
$H_{al}^{+}$, and these correlations will be related to the ones
coming from the black hole horizon. Both sets of correlations give
rise to correlations between $\cal I^+$ and the future black hole
horizons, which give the particle creation, so the particle creation
due to the acceleration horizon should just be the image under the
left-right interchange of the particle creation due to the black hole
horizon.  This justifies our only calculating the latter contribution.

We now calculate $\beta^{b}_{\omega \omega' lm}$. The modes $p_\omega,
  \bar{p}_\omega, q_\omega, \bar{q}_\omega$ are orthogonal, and
\begin{equation}
(p_\omega, p_{\omega'}) = 2 \tilde{\omega} \delta^3(\vec{k} - \vec{k}') =
\frac{2}{\tilde{\omega} \sin \theta_0} \delta(\eta_0 - \eta_0') \delta
(\theta_0 - \theta_0') \delta(\tilde{\omega} - \tilde{\omega}').
\end{equation}
Thus, we can use (\ref{bog1}) to show
\begin{equation}
\bar{\beta}^{b}_{\omega \omega' lm} =  \frac{\tilde{\omega} \sin
\theta_0}{2}(p_\omega,\bar{f}^{b}_{\omega' lm}) = \frac{2A
\omega}{|G'(\xi_4)|}(p_\omega,\bar{f}^{b}_{\omega' lm}) .
\end{equation}
To evaluate this inner product, we need to express both the modes as
functions on the same Cauchy surface. We do this by evolving the mode
$\bar{f}^{b}_{\omega' lm}$ forwards from $\cal C$ to $\tilde{\cal
C}$.

The propagation from $\cal C$ to $\tilde{\cal C}$ can be broken up
into two stages: propagation through the right Rindler diamond, from
$H_{al}^{-}$ and $H_{br}^{-}$ to $H_{ar}^{+}$ and $H_{bl}^{+}$, and
propagation through the future diamond, from $H_{al}^{+}$ and
$H_{ar}^{+}$ to ${\cal I}^+$. The initial data on $H_{br}^{-}$ is just
the restriction of (\ref{bhp1}) to the past part of the black hole
horizon, while $\bar{f}^{b}_{\omega' lm}$ vanishes on $H_{al}^{-}$.
From the discussion of (\ref{zeq}), we recall that at the acceleration
and black hole horizons, $\gamma(y) \sim e^{\pm i \omega z}$.  Using
this and the form (\ref{bhp1}) of the mode $f^{b}_{\omega' lm}$, we
find that the boundary conditions on $\gamma(y)$ are
\begin{equation} \label{bc1}
\gamma(y)  = e^{-i \omega z} + C_R e^{i\omega z}
\end{equation}
at the black hole horizon $z \to \infty$, and
\begin{equation} \label{bc2}
\gamma(y)  = C_T e^{-i \omega z} 
\end{equation}
at the acceleration horizon $z \to -\infty$, where $C_R$ and $C_T$ are
constants which remain to be determined. In appendix \ref{trans}, we
solve (\ref{fyeq}) with these boundary conditions in the limit $r_+A
\ll 1$, assuming $\omega \sim O(1)$, and find that $C_T \sim
(\alpha-1)^{2l+1}$, and that, for $l_0=0$, $|C_T| \approx (\alpha-1)
\omega/2$. Because the transmission factor $C_T$ is increasingly
suppressed for increasing $l$, we will be mostly interested in the
contribution from the $l_0=0$ mode, as the other contributions will be
smaller than the terms that we neglect in our approximate calculation
of the $l_0=0$ contribution.

The propagation from $H_{al}^{+}$ and $H_{ar}^{+}$ to ${\cal I}^+$ is
also described in appendix \ref{trans}. This part of the calculation
is substantially easier; it is very similar to solving the angular
equation (\ref{nxeq}).  In the conformal frame where the metric has
the form (\ref{scricm}), the restriction of $f^{b}_{\omega' lm}$ to
$\cal I^+$ is
\begin{equation}
f^{b}_{\omega' lm} |_{\cal I^+} = \frac{2N C_T G^{1/2}(y)}{
|1-e^{-2\pi \omega'/\kappa}|^{1/2} |G'(\xi_4)|\sin \theta} 
e^{-i \omega' (t+z) }
e^{i m \varphi} e^{-m |z|} f_{l \omega'}(p) \tilde{n}_{lm}(p). 
\end{equation}
In this expression, $f_{l \omega'}(p)$ is given by the definition at
the end of appendix \ref{trans}, and we have defined
$\tilde{n}_{lm}(p)$ to be $n_{lm}(p)$ for $p < 1/2$ ($\chi < 0$) and
$e^{2 m \chi} n_{lm}(p)$ for $p > 1/2$ ($\chi > 0$), where $n_{lm}(p)$
is the approximate solution of the angular equation given in appendix
\ref{ang}. When $p \to 0$, $f_{l \omega'} (p), \tilde{n}_{lm}(p) \to
1$. When $p \to 1$, $\tilde{n}_{lm} (p) \to e^{i \varrho}$, some
constant phase.

Evaluating the inner product, we find that
\begin{equation} \label{bhbog}
\bar{\beta}^{b}_{\omega \omega' lm} = - \frac{2\bar{N} \bar{C}_T
\omega e^{i \omega' t_0} e^{-i m\varphi_0}}{\sqrt{2\pi} |G'(\xi_4)|
|1-e^{-2\pi \omega'/\kappa}|^{1/2}} \int dz \, e^{i\omega r}
e^{i\omega'z} e^{-m|z|} f_{l \omega'}(p) \tilde{n}_{lm} (p),
\end{equation}
where $t=t_0$ corresponds to $\theta=\theta_0$, 
$\varphi=\varphi_0$ corresponds to $\eta=\eta_0$, and we have used $dr
= dz / G^{1/2}(y)$, which follows from (\ref{r}) and (\ref{zchi}).

Note that apart from an overall phase, this expression depends only on
the frequency $\omega$, and not on $\theta_0, \eta_0$. This means that
the expression is boost invariant, that is, invariant under
translations in $t$, as the orbits of the boosts are the
cross-sections labeled by $r$, and thus these boosts preserve the
frequency $\omega$ with respect to $r$.

We can't evaluate the integral in (\ref{bhbog}), but we can still get 
some interesting physical information about the radiation out of this
expression. Because $G(y) \to 0$ as $z \to \pm \infty$,
\begin{equation}
\frac{dr}{dz} = \frac{1}{G^{1/2}(y)} \to \pm \infty \mbox{ when } z
\to \pm \infty,
\end{equation}
and hence the $e^{i \omega r}$ part of the integrand oscillates with
an effective frequency which tends to infinity at large $|z|$. Since
the amplitude is bounded, the main contribution to the integrand will
come from the region near $z=0$ where the integrand oscillates slowly.

The integral in (\ref{bhbog}) will give an answer which is peaked in
$\omega'$ with some finite width, so the integration over $\omega'$ of
$|\bar{\beta}^{b}_{\omega \omega' lm}|^2$ in (\ref{numop}) should give
a finite answer. By contrast, in the case of a static black hole, the
analogous formula for the Bogoliubov coefficient gives a delta
function in $\omega'$, so the expected number of particles is infinite
(that is, in that case there is a steady flux of particles across
$\cal I^+$).

Our calculation of the transmission factor in appendix \ref{trans} is
only valid for $|\omega'| \leq 1$, and we might expect that for
sufficiently large $\omega'$, the potential barrier would become
unimportant, and $C_T \sim O(1)$. However, the Bogoliubov coefficient
will be small for large negative $\omega'$ because of the factor
$|1-e^{-2\pi \omega'/\kappa}|^{-1/2}$. We also expect that it would be
small at large positive $\omega'$, as the integrand in the integral in
(\ref{bhbog}) will then oscillate rapidly for all values of $z$, making
the integral small. Thus, the main contribution to the integral over
$\omega'$ in (\ref{numop}) will come from small negative $\omega'$,
where the calculation of $C_T$ is valid.

We expect that the size of the contribution from each $l,m$ will be
primarily determined by the transmission factor, so we expect that the
contribution from $l_0=m_0=0$ will dominate the summation over $l,m$
in (\ref{numop}). We now consider the form of this contribution in the
point-particle limit, where we can somewhat simplify the expressions
and illustrate some of these remarks.  When $(\alpha-1) \ll 1$, we
have $G(y) \approx 4 p(1-p)$ on $\cal I^+$, where $p =
(\hat{y}-1)/(\alpha-1)$. Further, $z \approx \frac{1}{2}
\ln(p/(1-p))$, as $0 \leq p \leq 1$ on $\cal I^+$, so
\begin{equation}
G(y) \approx \frac{1}{ \cosh^2 z}.
\end{equation}
Thus, $dr/dz \approx  \cosh z$, and hence
\begin{equation}
r \approx  \sinh z.
\end{equation}
Also, $\kappa \approx 1, f_{0 \omega}(p) \approx 1$, and
$\tilde{n}_{00}(p) \approx 1$. Therefore
\begin{equation} \label{bogapp}
\bar{\beta}^{b}_{\omega \omega' 00} \approx - \frac{\bar{N}
\bar{C}_T \omega
e^{i \omega' t_0} e^{-i m\varphi_0}}{\sqrt{2\pi} 
|1-e^{-2\pi \omega'}|^{1/2}} \int dz \, e^{i (\omega' z +
 \omega \sinh z)}. 
\end{equation}
As we argued above, the main contribution to the integration will come
from the region near $z=0$, so the primary contribution to
$\bar{\beta}^{b}_{\omega \omega' 00}$, and hence to the number
operator, will come from the part of the generator closest to the
points where the black holes intersect $\cal I^+$. If we restrict our
attention to the region near $z=0$, we can expand $\sinh z$ in a power
series, and we see that the integrand is most nearly constant near
$z=0$ if $\omega' = -\omega$, so we expect that the Bogoliubov
coefficient will be peaked at $\omega' = -\omega$. This peak will
become narrower as $\omega \to 0$, approaching a delta function in the
limit, but the amplitude tends to zero in this limit because of the
factor of $\omega$ in front of the integral, so this does not imply
infinite particle production.

The leading-order part of the total particle production along the
generator labeled by $\theta_0, \eta_0$ is given by integrating
$|\beta^{b}_{\omega \omega' 00}|^2$ over $\omega$ and $\omega'$; we
can't do this integral, but given the arguments above, it seems
reasonable to expect the answer to be finite. The integration over
all generators, which gives the total particle production, will not
give rise to any divergences either. 

\section{Discussion}

In the first part of this paper, we argued that the scattering off
virtual black hole pairs, which could lead to loss of quantum
coherence in ordinary scattering processes, could be discussed in
terms of a path integral over Euclidean metrics with topology $S^2
\times S^2 - \{\mbox{point}\}$.  In this approach, one considers the
scattering in each metric and performs a path integral over all such
metrics. Since we cannot perform this path integral, we then
restricted the discussion to one such metric, and analytically
continued the solution to a Lorentzian section to make the scattering
easier to understand.

We argued that the appropriate quantum state is the
analytically-continued Euclidean vacuum state $|0\rangle_E$, and
we argued that this state will contain a finite, non-zero number of
particles at infinity. It is well-known that from the point of view of
an observer co-moving with the black holes, this state corresponds to
a thermal equilibrium between the black holes and a thermal bath of
acceleration radiation.  Thus, this state must be time-reversal
invariant, which means that the particle content at past null infinity
$\cal I^-$ is the time-reverse of the particle content at future null
infinity $\cal I^+$. This implies that no net energy is gained or lost
by the black holes in this scattering process, which is what we would
expect for a model of a virtual loop, and is in agreement with the
fact that the state is an equilibrium as seen by co-moving observers.

The fact that there is a non-zero number of particles at $\cal I^+$
implies that there is loss of quantum coherence in this semi-classical
calculation, as each particle detected at infinity can be thought of
as one member of a virtual pair, the other one of which has fallen
into the black hole, carrying away information. More formally, there
are correlations between modes on future infinity and modes on the
future black hole horizon, and the information encoded in these
correlations is lost because we do not observe the state on the future
black hole horizon.  This loss of quantum coherence is of the same
character as that observed in static black holes.

In the second part of the paper, we proceeded to an explicit
calculation of the scattering in the $C$ metric. Although the
Euclidean $C$ metric solution has topology $S^2 \times S^2 -
\{\mbox{point}\}$, it is not usually thought of as describing a
virtual black hole loop, as it is a solution of the field equations,
and it has a conical singularity along one of the axes. However, we
believe it is a reasonably good model for a virtual black hole loop,
and the wave equation separates in this background, so it is
relatively easy to study the scattering explicitly. The $C$ metric is
asymptotically flat \cite{ash:81}, so it is also straightforward to
study the radiation at infinity. One slightly surprising fact about
the structure at infinity is that the affine parameter along
generators of $\cal I^+$ is $\tilde{r}$, which is spacelike between
the black hole and acceleration horizons, while the boost time
coordinate $t$ becomes a spacelike coordinate labeling the generators
of $\cal I^+$.

It is also worth noting that the transmission factor $C_T \sim
(\alpha-1)^{2l+1}$. This implies that the dominant contribution to the
particle production is in the s-wave, as for static black holes,
because of the high centrifugal potential barrier for higher-spin
modes. It also suggests that the scattering of higher-spin fields
off such virtual black hole loops will be suppressed relative to that
of scalar fields, as they cannot radiate in the s-wave. This is in
agreement with the arguments of \cite{hawk:bub1,hawk:bub2}. 

The calculation we have actually been able to perform is rather
limited; we considered only one specific, rather special metric, and
we were only able to study the scattering on it in a particular limit.
However, the results we have obtained give an estimate of the
magnitude and nature of the effects of virtual black hole loops, and
they agree well with our general expectations.

\section{Acknowledgements}

We thank Gary Horowitz and Don Marolf for useful discussions. SFR
thanks St. John's College Cambridge and the Natural Sciences and
Engineering Research Council of Canada for financial support. SFR's
work was also supported in part by NSF Grant PHY95-07065.

\appendix

\section{The angular quantization condition}
\label{ang}

In the point-particle limit $(\alpha-1) \ll 1$, the deviations from
spherical symmetry in the $x,\varphi$ part of the metric become small,
so we would expect that the dependence on $x$ will reduce to the usual
angular momentum modes, with quantum numbers $l$ and $m$. Recall that
because of the periodicity of $\varphi$, $m= m_0 |G'(\xi_4)|/2 =
m_0 [1 + O(\alpha-1)]$, where $m_0$ is an integer. We also
expand $l = l_0 + l_1 (\alpha-1) + \ldots$. The range of $\hat{x}$
is $1 \leq \hat{x} \leq \alpha$, so we define a new coordinate $p =
(\hat{x}-1)/(\alpha-1)$. If we expand $\hat{n}(\hat{x})$ in powers of
$\alpha-1$, $\hat{n}(\hat{x}) = n_{lm}(p) = n_{0}(p) +
(\alpha-1) n_{1}(p) + \ldots$, then (\ref{nxeq}) can be separated into a
series of equations for these functions. The first equation is
\begin{equation}
\partial_p [p(p-1) \partial_p n_{0}(p)] - m_0
\partial_p n_{0}(p) - 
l_0 (l_0+1) n_{0}(p)= 0.
  \label{peq}
\end{equation}

This equation is a hypergeometric equation. The possible values of
$l_0$ are restricted by requiring that the solution behave
appropriately at the two poles, $p=0,1$. As we said earlier, for the
solution for $\phi$ to be physically relevant, we must have $\nu(x)
\sim e^{-m |\chi|}$ as $\chi \to \pm \infty$.  That is, we require
that $\nu(x)$, and hence $\phi$, doesn't blow up at the axes. Since
$\hat{\nu}(\hat{x}) = e^{m \chi} \hat{n}(\hat{x})$, the appropriate
boundary conditions on $n_{lm}(p)$ are that $n_{lm}(p) = 1$ as $\chi
\to -\infty$, which corresponds to $p=0$, and $n_{lm}(p) \sim e^{-2m
\chi}$ as $\chi \to \infty$, which corresponds to $p=1$. Therefore,
the appropriate solution of the hypergeometric equation (\ref{peq}) is
$n_{0}(p) = F(l_0+1,-l_0;1+m_0;p)$, where $F$ is the hypergeometric
series, as $F(a,b;c;p) \to 1$ as $p \to 0$. If we analytically
continue this solution to a neighborhood of $p=1$, we find
\begin{eqnarray}
  n_{0}(p) &=& \frac{\Gamma(1+m_0) \Gamma(m_0)}{\Gamma(m_0-l_0)
\Gamma(m_0+l_0+1)} F(l_0+1,-l_0,1-m_0; 1-p) \nonumber \\ &&+
\frac{\Gamma(1+m_0) 
\Gamma(-m_0)}{\Gamma(l_0+1) \Gamma(-l_0)} (1-p)^{m_0}
F(m_0-l_0,1+m_0+l_0;1+m_0;1-p). 
\end{eqnarray}
The second term has the appropriate behavior for $p\to 1$, since
$e^{-\chi} \approx (1-p)^{1/2}$ for $p \approx 1$. Thus, the
coefficient of the first term must vanish, which can only happen if
$l_0-m_0$ is a non-negative integer.  This is just the usual
quantisation condition for angular momentum, and $l_0$ is thus the
total angular momentum quantum number.

The next-order term $l_1$ can similarly be fixed by requiring that
the solution $n_1(p)$ is regular at $p=0,1$.  Unfortunately, it is not
possible to give a general formula for $l_1$; the equation must be
solved separately for each $l_0,m_0$. We are particularly interested
in the case $l_0=m_0=0$, as we expect this mode to make the dominant
contribution to the particle production on $\cal I^+$. In this case,
$n_{0}(p) = F(1,0;1;p)= 1$, while the equation for $n_1(p)$ is
\begin{equation}
\partial_p [p(p-1) \partial_p n_{1}(p)] = l_1-p.
\end{equation}
This equation has a solution which is regular at $p=0,1$ only if $l_1
= 1/2$; in this case, the solution is $n_{1}(p) = -p/2 + C$, where $C$
is a constant. One can similarly fix all the $l_i$. 

\section{The transmission factor}
\label{trans}

In section \ref{bog}, we found that to evolve the positive-frequency
modes from $\cal C$ to $\tilde{\cal C}$, we need to calculate the
transmission factor $C_T$ between the black hole and acceleration
horizons. That is, we need to solve (\ref{fyeq}) with the boundary
conditions (\ref{bc1},\ref{bc2}), and find $C_T$. For convenience, we
will repeat those here. The equation is 
\begin{equation} \label{fyeqr}
\partial_{\hat{y}}[(\hat{y}^2 -1)(\hat{y}^2-\alpha^2)
\partial_{\hat{y}} \hat{f}(\hat{y})] +2i\omega \zeta
\partial_{\hat{y}} \hat{f}(\hat{y}) + 
2(\hat{y}^2 - \beta_D^2) \hat{f}(\hat{y}) = 0.
\end{equation}
In terms of the function $\hat{f}(\hat{y})$, the boundary conditions
are
\begin{equation}  \label{fbc1}
\hat{f}(\hat{y}) = 1 + C_R e^{2i \omega z}
\end{equation}
near the black hole horizon $\hat{y} = -1$ and 
\begin{equation} \label{fbc2}
\hat{f}(\hat{y}) = C_T
\end{equation}
near the acceleration horizon $\hat{y}=1$. 

We can't solve this equation exactly, but if $(\alpha-1) \ll 1$, then
we can solve it approximately. First note that if $\hat{y}^2-1$ is
$O(1)$ (that is, if $\hat{y}$ is not close to $\pm 1$), we can 
neglect terms involving $\alpha-1$ to approximate (\ref{fyeqr}) as
\begin{equation}
\partial_{\hat{y}}[(\hat{y}^2 -1)^2
\partial_{\hat{y}} \hat{f}(\hat{y})] +
2(\hat{y}^2 - \beta_D^2) \hat{f}(\hat{y}) = 0.
\end{equation}
In neglecting the term involving $\omega$, we have made the further
assumption that $|\omega| \sim O(1)$; that is, that $\omega$ is not
large. This equation is now a hypergeometric equation. To put it in
the standard form, we set $\hat{f}(\hat{y}) = 2^a (1-\hat{y}^2)^{-a}
(\alpha-1)^a g(s)$, where $s = (\hat{y}+1)/2$, $a=l+1$. Then
\begin{equation} \label{ap1}
s(s-1) \partial_s^2 g(s) - 2l(2s-1)\partial_s g(s) + 2l(2l+1)g(s) = 0,
\end{equation}
where we have used $\beta_D = 1 + 2l(l+1)$. We use $l$ rather than
$l_0$ in the approximate equations in this section, because regarding
$l$ as an integer would introduce degeneracies in the approximate
equations which are not present in the exact equation. Near
$\hat{y}=\pm 1$, the solutions of (\ref{ap1}) can be expressed in
terms of hypergeometric series about $\hat{y}=\pm 1$.  However, we
cannot approximate (\ref{fyeqr}) by (\ref{ap1}) in a neighborhood of
radius $O(\alpha-1)$ around $\hat{y} = \pm1$, which is precisely where
we wish to impose boundary conditions.

Therefore we need a separate approximation to cover these
neighborhoods. When $\hat{y}^2 -1 \sim (\alpha-1)$, make a coordinate
transformation $\hat{y} = \pm (1 + (\alpha-1) q_\pm)$. Then if we keep
just the leading terms, (\ref{fyeqr}) becomes
\begin{equation}
\partial_{q_\pm} [q_\pm(q_\pm-1) \partial_{q_\pm} f(q_{\pm})] \pm i
\omega \partial_{q_\pm} f - l(l+1) f=0, 
\end{equation}
where $f(q_\pm) = \hat{f}(\hat{y})$. These are, once again,
hypergeometric equations. The solution about $\hat{y} = -1$ which
satisfies the boundary condition (\ref{fbc1}) is
\begin{equation} \label{qm}
f(q_-) = F(a,b;2-c;q_-) + C_R (-q_-)^{-i \omega} F(b+c-1,a+c-1;c;q_-),
\end{equation}
and the solution about $\hat{y}=1$ which satisfies the boundary
condition (\ref{fbc2}) is
\begin{equation} \label{qp}
f(q_+) = C_T F(a,b;c;q_+),
\end{equation}
where $F$ is the hypergeometric function, $a=l+1, b=-l$ and
$c=1-i\omega$. Now analytically extend these solutions to large
$q_\pm$: at large $q_-$, the solution (\ref{qm}) becomes
\begin{eqnarray} \label{ld1}
f(q_-) &=& \frac{\Gamma(c) \Gamma(b-a)}{\Gamma(c-a) \Gamma(b)} \left(
C_R + \frac{\Gamma(2-c)\Gamma(c-a)}{\Gamma(c) \Gamma(2-c-a)} \right)
(-q_-)^{-a} \\ \nonumber
&&+ \frac{\Gamma(c) \Gamma(a-b)}{\Gamma(c-b) \Gamma(a)} \left(
C_R + \frac{\Gamma(2-c)\Gamma(c-b)}{\Gamma(c) \Gamma(2-c-b)} \right)
(-q_-)^{-b}, 
\end{eqnarray}
while at large $q_+$,  the solution (\ref{qp}) becomes
\begin{equation} \label{ld2}
f(q_+) = C_T \frac{\Gamma(c) \Gamma(b-a)}{\Gamma(c-a) \Gamma(b)}
(-q_+)^{-a}+  C_T \frac{\Gamma(c) \Gamma(a-b)}{\Gamma(c-b) \Gamma(a)}
(-q_+)^{-b}.
\end{equation}

Now for $1 \ll |q_\pm| \ll (\alpha-1)^{-1}$, both approximations are
applicable, so we can use the large-distance behavior
(\ref{ld1},\ref{ld2}) of the approximation for $\hat{y}$ near $\pm 1$
as boundary data for the approximation (\ref{ap1}). If we pick the
solution $g(s)$ to be
\begin{eqnarray}
g(s) &=& \frac{\Gamma(c) \Gamma(b-a)}{\Gamma(c-a) \Gamma(b)} \left(
C_R + \frac{\Gamma(2-c)\Gamma(c-a)}{\Gamma(c) \Gamma(2-c-a)} \right)
F(-2l,-2l-1;-2l;s) \\ \nonumber
&&+ \frac{\Gamma(c) \Gamma(a-b)}{\Gamma(c-b) \Gamma(a)} \left(
C_R + \frac{\Gamma(2-c)\Gamma(c-b)}{\Gamma(c) \Gamma(2-c-b)} \right)
(\alpha-1)^{b-a} 2^{a-b}
s^{a-b} F(0,1;2l+2;s),
\end{eqnarray}
then the boundary conditions obtained from (\ref{ld1}) are
automatically satisfied. We can analytically continue this solution to
a neighborhood of $s=1$; to satisfy the boundary conditions obtained
from (\ref{ld2}) in this neighborhood at the same time, we must
require
\begin{equation}
C_T \frac{\Gamma(c) \Gamma(b-a)}{\Gamma(c-a) \Gamma(b)} =
\frac{\Gamma(c) \Gamma(a-b)}{\Gamma(c-b) \Gamma(a)} \left( 
C_R + \frac{\Gamma(2-c)\Gamma(c-b)}{\Gamma(c) \Gamma(2-c-b)} \right)
(\alpha-1)^{b-a} 2^{a-b}
\end{equation}
and
\begin{equation}
C_T \frac{\Gamma(c) \Gamma(a-b)}{\Gamma(c-b) \Gamma(a)}
(\alpha-1)^{b-a} 2^{a-b} = \frac{\Gamma(c) \Gamma(b-a)}{\Gamma(c-a)
\Gamma(b)} \left( C_R + \frac{\Gamma(2-c)\Gamma(c-a)}{\Gamma(c)
\Gamma(2-c-a)} \right).
\end{equation}

Solving these two equations for $C_R$ and $C_T$, we find
\begin{equation}
C_T = -e^{i \vartheta} \frac{\delta - \bar{\delta}}{1-\delta^2}
\end{equation}
and
\begin{equation}
C_R = -e^{i \vartheta} + \delta C_T,
\end{equation}
where 
\begin{equation}
e^{i \vartheta} = \frac{\Gamma(2-c) \Gamma(c-b)}{\Gamma(c) \Gamma(2-c-b)}
\end{equation}
and
\begin{equation}
\delta = \left( \frac{\alpha-1}{2} \right)^{a-b} \frac{\Gamma(b-a)
\Gamma(a) \Gamma(c-b)}{\Gamma(a-b) \Gamma(b) \Gamma(c-a)}.
\end{equation}
Note that these coefficients satisfy $|C_T|^2 + |C_R|^2 =1$, as they
should. 

After some manipulation, we find 
\begin{equation}
\delta - \bar{\delta} = -\frac{4i}{2l+1} \left( \frac{\alpha-1}{8}
\right)^{2l+1} \frac{\Gamma(1+l-i\omega)
\Gamma(1+l+i\omega)}{\Gamma(l+\frac{1}{2})^2} \sinh \pi \omega.
\end{equation}
Also, $\delta \sim (\alpha -1)^{2l+1}$, so the denominator in $C_T$
can be ignored for this leading-order calculation. For large $l$, we
thus find
\begin{equation}
C_T \approx 2 e^{i (\vartheta + \frac{\pi}{2})} \left( \frac{\alpha-1}{8}
\right)^{2l+1} \sinh \pi \omega,
\end{equation}
while for $l_0=0$, we find
\begin{equation}
C_T \approx  e^{i (\vartheta
 + \frac{\pi}{2})} \left( \frac{\alpha-1}{2}
\right) \omega.
\end{equation}
These results are valid for $(\alpha-1) \ll 1$ and $|\omega| \leq 1$. 

We have found the value of $\hat{f}(\hat{y})$ at the acceleration
horizon $\hat{y}=1$. The region between $H_{al}^{+}, H_{ar}^{+}$, and
${\cal I}^+$ is the region between $\hat{y}=1$ and $\hat{y}=\hat{x}$;
to evolve $\hat{f}(\hat{y})$ through this region, we just need to find
the form of $\hat{f}(\hat{y})$ between $\hat{y}=1$ and
$\hat{y}=\alpha$, which will also be the solution on $\cal I^+$. Now,
the approximation (\ref{qp}) is valid throughout this region, so the
result is simply that on $\cal I^+$, 
\begin{equation}
\hat{f}(\hat{y}) = C_T f_{l \omega}(p) \approx C_T F(a,b;c;p), 
\end{equation}
where $a,b,c$ are as in (\ref{qp}). Note that $\hat{x} = \hat{y}$
implies $q_+ = p$. For $l_0=0$, the leading-order part of this
solution is $f_{0\omega}(p) \approx 1$, just as for $n_{lm}(p)$. 

\begingroup\raggedright\endgroup

\end{document}